\documentstyle[12pt]{article}

\newcommand{\lsim}{\mathrel{\lower4pt\hbox{$\sim$}}
\hskip-12.5pt\raise1.6pt\hbox{$<$}\;}

\newcommand{\gsim}{\mathrel{\lower4pt\hbox{$\sim$}}
\hskip-12.5pt\raise1.6pt\hbox{$>$}\;}

\begin{document}
\baselineskip 18pt plus 2pt

\noindent \hspace*{13cm}UCRHEP-T198\\
\noindent \hspace*{13cm}July 1997\\

\begin{center}
{\bf TWO HIGGS DOUBLETS MODELS AND CP VIOLATING HIGGS EXCHANGE IN 
$e^+e^- \to t \bar t Z$}
\vspace{.7in}

S. Bar-Shalom$^a$, D. Atwood$^b$ and A. Soni$^c$\\
\end{center}
\vspace{.2in}

\noindent
$^a$ Physics Dept., University of California, Riverside CA 92521, USA.\\
$^b$ Theory Group, Thomas Jefferson National Accelerator Facility, Newport News, VA 23606, USA.\\
$^c$ Physics Dept., Brookhaven Nat.\ Lab., Upton NY 11973, USA.
\vspace{0.5in}

\begin{center}
{\bf Abstract}\\
\end{center}
Appreciable CP Asymmetries ($\sim 10\%$) can arise in the reaction  
$e^+e^- \to t \bar t Z$ already at {\it tree-level} in models with two Higgs doublets. For a neutral Higgs particle, $h$, with a mass in the range  
50 GeV${} \lsim m_h \lsim 400$ GeV, it may be possible to 
detect a 2--3 sigma CP-odd effect in $e^+e^- \to t \bar t Z$ 
in $\sim 1-2$ years of running of a future high 
%ASf
energy $e^+e^-$ collider with c.m.\ energies of $\sim 1-2$ TeV and an integrated luminosity of 200-500 inverse fb.

\pagebreak

A future high energy $e^+e^-$ collider running at c.m.\ energies of 0.5--2
TeV,  often referred to as the Next Linear Collider (NLC),  will no doubt
serve as a very useful laboratory for a detailed  study of the properties
of the Higgs particle(s) and that of the top quark \cite{nlc}.  In particular,
it may  unveil new phenomena, beyond the Standard Model (SM) associated 
with the top Yukawa couplings to scalar particle(s). Evidence of  such new
$ttH$ couplings, if detected at the NLC, can give us important clues about
the nature of the scalar potential and of the properties of the
scalar particle(s). 

In the SM, the scalar potential is economically composed of only one
scalar doublet. Even a mild extension of the SM  with an additional scalar
doublet \cite{gunion}, can give  rise to rich new phenomena beyond 
the SM associated with top-Higgs systems, e.g.,  tree-level CP-violation
\cite{cp2hdm,atwood} and tree-level  flavor-changing-scalar (FCS) transitions
\cite{fc2hdm}, in interaction of  neutral scalars with the top quark.  
Indeed, the top quark, being so heavy, $m_t \sim 175$ GeV,    
is the most sensitive to these new interactions.

In this Letter we explore the possibility of detecting {\it tree-level} CP-violation
in the reaction  $e^+e^- \to t \bar t Z$. We find that in the best cases 
one needs about one thousand $t \bar t Z$ events to observe a 3-sigma
CP-nonconserving signal, which, as we will show here, may well be within
the experimental reach of the NLC\null.  To some extent, our findings for
$e^+e^- \to t \bar t Z$ are somewhat similar to our previous study of $e^+e^-
\to t \bar t h$ where large tree-level CP violation was reported \cite{cp2hdm}.
The process $e^+e^- \to t \bar t Z$ provides another independent, but analogously,
promising    venue to search for the signatures of the same CP-odd phase,
residing in the top-neutral Higgs  coupling, in future experiments. 
    
In the presence of two Higgs doublets the most general Yukawa 
lagrangian can be written as:

\begin{equation}
{\cal L}_Y = U^1_{ij} {\bar q}_{i,L} {\tilde {\phi}}_1 u_{j,R} + 
D^1_{ij} {\bar q}_{i,L} {\phi}_1 d_{j,R} + 
U^2_{ij} {\bar q}_{i,L} {\tilde {\phi}}_2 u_{j,R} + 
D^2_{ij} {\bar q}_{i,L} {\phi}_2 d_{j,R} + {\rm h.c.} \label{yukawa}~,
\end{equation}

\noindent where $\phi_i$ for $i=1,2$ are the two scalar doublets and $U^k_{ij},D^k_{ij}$,
for $k=1,2$, are the Yukawa  couplings matrices which are in general non-diagonal.
Depending  on the assumptions made, one can then obtain different versions
of a Two Higgs Doublet Model (2HDM).  In particular, if one imposes the
 discrete symmetries $\phi_1;\phi_2 \to -\phi_1;\phi_2$  and $d_{i,R};u_{i,R}
 \to -d_{i,R};-u_{i,R}~or~-d_{i,R};u_{i,R}$  one arrives at the so called
 Model~I or Model~II, respectively,  depending on whether the -1/3 and 2/3
 charged quarks are coupled  to the same or to different scalar doublets.
 If, in addition,  these discrete symmetries are softly violated by a  mass-dimension-two
 term in the Higgs potential,  then the real  and imaginary parts of the
 Higgs doublets mix, giving rise to  CP-violating scalar-pseudoscalar mixed
 couplings of a neutral  Higgs to fermions already at the tree-level
 \cite{froggat}. On the other hand, if one does not impose the above discrete
 symmetries, one arrives at a most general version of the 2HDM, often called
 Model~III, in which both FCS  transitions and CP-nonconserving interactions
 between the neutral  Higgs particles and fermions are present at tree-level
 (see e.g., Luke and Savage in \cite{fc2hdm} and \cite{sonirev}).   
 
The scalar spectrum of any of the above 2HDM's consists of three neutral 
Higgs and two charged Higgs particles. The ${\cal H}^kq \bar q$ and 
${\cal H}^k ZZ$ ($k=1,2,3$ corresponding to the three neutral Higgs particles
${\cal H}^k$  and $q$ stands for quark) interaction lagrangian parts of 
a  general 2HDM can be written as:  

\begin{equation}
{\cal L}_{{\cal H}^k qq}= -\frac{g_W}{\sqrt 2} \frac{m_q}{m_W} {\cal H}^k 
\bar q \left( a_q^k + i b_q^k \gamma_5 \right) q ~~,~~
{\cal L}_{{\cal H}^k ZZ}= \frac{g_W}{c_W} m_Z c^k 
{\cal H}^k g_{\mu \nu} Z^{\mu} Z^{\nu} \label{hqqhzz}~.  
\end{equation}

\noindent Note that in the SM the couplings in Eq.~\ref{hqqhzz},  of
the only neutral Higgs present, are $a_q=1/\sqrt {2},b_q=0$  and $c=1$ and
there is no phase in the ${\cal H}^kq \bar q$ coupling. In Model~II, for 
up quarks for example \cite{froggat}:

\begin{equation}
a_u^k=R_{1k}/\sin\beta~~,~~b_u^k=R_{3k}/\tan\beta~~,~~c^k=R_{1k}\sin\beta
+R_{2k}\cos\beta \label{mod2coup} ~,
\end{equation}

\noindent where $\tan\beta \equiv v_u / v_d$ and $v_u$($v_d$) is the vacuum-expectation-value
responsible for giving mass to  the up(down) quark. $R$ is the neutral Higgs
mixing  matrix which can be parameterized by three Euler angles  $\alpha_{1,2,3}$
\cite{froggat}.  

A general feature of the above 2HDM's is that only two out of the three neutral Higgs can simultaneously have a coupling to vector bosons and a 
pseudoscalar coupling to fermions. We will denote these two neutral Higgs
by $h$ and $H$ with couplings $a_q^h,b_q^h,c^h$ and  $a_q^H,b_q^H,c^H$, corresponding
to the light  and heavy neutral Higgs, respectively.\footnote{In some instances
we will denote these two neutral Higgs  by ${\cal H}$. Then ${\cal H}=h~or~H$
is to be understood.}  Then, an important aspect of these 2HDM's,  which
has crucial phenomenological implications for CP-violation, is  
that these couplings are subject to the constraint $b_q^hc^h + b_q^Hc^H=0$
\cite{atwood}. This implies the existence of a ``GIM''-like cancellation, 
namely; all CP-violating effects due to the Higgs sector, being proportional
to $b^hc^h + b^Hc^H$, must vanish when  the two Higgs states $h$ and $H$
are degenerate.         

We now discuss the possibility of having CP-violation, {\it already at the
tree-level},  driven by 2HDM's, in our reaction: 

\begin{equation}
e^+(p_+)+e^-(p_-)\to q(p_q) + 
\bar{q}(p_{\bar{q}}) + Z(p_Z) \label{eettz} ~.
\end{equation}

\noindent In the unitary gauge the reaction in Eq.~\ref{eettz} can proceed
via the Feynman diagrams depicted in Fig.~1.  Diagram $b$, where a pair of
$Z {\cal H}$  is produced (${\cal H}$ is produced either as real or virtual,
i.e.\ $m_{\cal H} >2m_t$ or $m_{\cal H}<2m_t$ respectively) followed by ${\cal
H} \to t \bar t$, is the only  one where new CP-nonconserving dynamics from
the Higgs  sector can arise being proportional to the  CP-odd phase in the
${\cal H} q \bar q$ vertex.  In particular, all CP-violating effects 
arise from the interference of diagram $b$ with the diagrams of class $a$
in Fig.~1 and  are proportional to the quantity $b_q^{\cal H} \times c^{\cal
H}$.   

A detailed cross-section analysis of the reaction $e^+e^- \to t \bar t Z$
%ASf within ..in
was performed in the SM by Hagiwara {\it et al}.\ in \cite{hagiwara}.
There, it was found that the Higgs exchange contribution of diagram $b$ in
Fig.~1 will be almost invisible in a TeV $e^+e^-$ collider for neutral Higgs
masses in the range $m_h < 2m_t$. On the contrary, we will show here that
if the scalar sector is doubled, then the lightest neutral Higgs may reveal
itself through CP-violating interactions with the top quark even if 
$m_h <2m_t$. We will sketch below the main characteristics of the total differential
cross-section (DCS) and focus primarily on its CP-violating part. The tree-level
polarized DCS,  $\Sigma^0_{(j)}$, $j=-1(1)$ for left(right) handed electrons,
is in general a sum of two terms: the CP-even and odd terms $\Sigma^0_{+(j)}$
and $\Sigma_{-(j)}^0$, respectively,  i.e.\ $\Sigma^0_{(j)} \equiv \Sigma_{+(j)}^0
+ \Sigma^0_{-(j)}$. However, we can furthermore divide $\Sigma^0_{\pm (j)}$
into: 

\begin{eqnarray}
&&\Sigma_{+(j)}^0 \equiv  \Sigma_{++(j)}^{0(SM)} +  \Sigma_{++(j)}^{0({\cal
H})} +  \Sigma_{+-(j)}^{0({\cal H})} \label{cpeven}~,\\
&&\Sigma_{-(j)}^0 \equiv \Sigma_{-+(j)}^{0({\cal H})} + \Sigma_{--(j)}^{0({\cal
H})} \label{cpodd}~, 
\end{eqnarray}

\noindent where the first and second subscripts denote the CP property and
the $T_N$ property ($T_N$ is the naive  time reversal operator defined by
replacing time with its negative without switching initial and final 
states) of the DCS's in Eqs.~\ref{cpeven} and \ref{cpodd}, respectively. The 
superscript indicates if it is a pure SM contribution, coming from diagrams
$a$ and denoted by (SM), or   interference terms associated with diagram
$b$ and  denoted by $({\cal H})$. Thus, for example, $\Sigma_{--(j)}^{0({\cal
H})}$ is the CP-odd, $T_N$-odd polarized  DCS, upon which we will concentrate,
that emanates from the interference of  diagram $b$ with the SM diagrams
$a$ in Fig.~1.  

It is then very simple to identify each of the DCS's in  Eqs.~\ref{cpeven}
and \ref{cpodd} associated with  the 2HDM-SM and the 2HDM-2HDM interferences
in terms  of the Higgs coupling constants $a_q^{\cal H},b_q^{\cal H}$ and
$c^{\cal H}$ defined in Eq.~\ref{hqqhzz}. In particular we find:

\begin{eqnarray} 
\Sigma_{++(j)}^{0({\cal H})} & = &  a^h_q c^h {\rm Re} (\Pi_h)f_{++(j)}^1 + 
(a^h_q c^h )^2 \left( {\rm Re} (\Pi_h)f_{++(j)}^2 +
{\rm Im} (\Pi_h) f_{++(j)}^3 \right) \nonumber \\
&&~+( b^h_q c^h)^2  \left( {\rm Re} (\Pi_h)f_{++(j)}^4 +
{\rm Im} (\Pi_h) f_{++(j)}^5 \right) + ~(h \to H) \label{sigpp}~,\\
\Sigma_{+-(j)}^{0({\cal H})} & = & a^h_q c^h {\rm Im} (\Pi_h) f_{+-(j)}^1 +
~(h \to H) ~,\\ 
\Sigma_{-+(j)}^{0({\cal H})} & = & b^h_q c^h {\rm Im} (\Pi_h) f_{-+(j)}^1 +
~(h \to H)  ~,\\  
\Sigma_{--(j)}^{0({\cal H})} & = & b^h_q c^h
{\rm Re} (\Pi_h) f_{--(j)}^1 + ~(h \to H) \label{sigmm}~,
\end{eqnarray}

\noindent where: 

\begin{equation}
\Pi_{\cal H} \equiv \left(s+m_Z^2-m_{\cal H}^2-
2p \cdot p_Z +i m_{\cal H} \Gamma_{\cal H} \right)^{-1} ~.
\end{equation} 

\noindent $p \equiv p_-+p_+$ and $\Gamma_{\cal H}$  is the width of ${\cal
H}$. $f_{mn (j)}^{\ell}$, $m,n=+~or~-$, are  kinematical functions of phase
space which transform like $m$ under CP and like $n$ under $T_N$. 
A few important remarks are in order at this stage:\\
1. The functions $f_{--(j)}^1$ and $f_{+-(j)}^1$, being $T_N$-odd, are 
proportional to the Levi-Civita tensor $\epsilon(p_-,p_+,p_q,p_{\bar q})$.\\    
2. There is no term proportional to $a^{\cal H}_q b^{\cal H}_q$ in the DCS's
at tree-level.\\ 
3. The diagrams where the $Z$ is emitted from the incoming electron and
positron  lines, do not contribute to $\Sigma_{--(j)}^{0({\cal H})}$.\\
4. $\Sigma_{+-(j)}^{0({\cal H})}$ and $\Sigma_{-+(j)}^{0({\cal H})}$ are
proportional  to the absorptive part coming from the Higgs propagator, ${\rm
Im} (\Pi_h)$, and are therefore not pure tree-level quantities being 
proportional to the Higgs width. Thus, a consistent calculation of the $+-$
and $-+$ parts of the DCS has to include  the full next order (i.e., 1-loop
order) contribution in perturbation theory. {\it In contrast}, $\Sigma_{--(j)}^{0({\cal
H})}$, being an odd  function of $T_N$, is proportional to ${\rm Re} (\Pi_h)$
and  is therefore a pure tree-level quantity.   

We will concentrate here on the CP-odd $T_N$-odd effects emanating from 
$\Sigma_{--(j)}^{0({\cal H})}$ which, as mentioned above, is proportional
at the tree-level to  the interference of the SM-like diagrams where the
$Z$ is radiated off the $t$ or $\bar t$ with the Higgs exchange diagram
\cite{oneloop}. By measuring a CP-odd $T_N$-odd observable one can extract
information on the magnitude of  $b_q^{\cal H} c^{\cal H}$. However, note
that the full  DCS contains more information about the other scalar  coupling
combinations. Thus with the appropriate optimal  observables \cite{sonioptimal},
it is possible   to isolate each coupling combination in Eqs.~\ref{sigpp}-\ref{sigmm},
and therefore, in principle, to  identify  the exact quantum numbers of the
neutral Higgs  particle, say the lightest one,  which is exchanged in diagram
$b$. This technique was applied  to the reaction $e^-e^+ \to t \bar t h$ 
by Gunion {\it et al.} in \cite{gunion1}. However, we expect that the reaction
$e^-e^+ \to t \bar t Z$ will be less sensitive to the CP-even top-Higgs couplings
due to the dominating presence of the SM diagrams $a$ depicted in Fig.~1.

%The expression for $\Sigma^0_{+(j)}$ is quite involved and 
%will not be given explicitly here. We will only write 
The CP-odd $T_N$-odd kinematic function of phase space, $f^1_{--(j)}$,
corresponding to $\Sigma_{--(j)}^{0({\cal H})}$,  and of relevance to the
present analysis is: 

\begin{eqnarray}
f_{--(j)}^1 &=& - \sqrt 2 \left( \frac{2g_W^3}{c_W^3} \right)^2 
\frac{m_q^2}{m_Z^2} \Pi_Z T_q^3 c_j^Z \epsilon(p_-,p_+,p_q,p_{\bar q}) \times
\nonumber\\ 
&& \left\{ j (\Pi_q+\Pi_{\bar q}) \left[ m_Z^2 w_j^- + (s_t-s) w_j^+ \right]
+   T_q^3 c_j^Z  \Pi_Z (\Pi_q-\Pi_{\bar q}) f \right\}~,
\end{eqnarray}

\noindent where:

\begin{equation}
w_j^{\pm} \equiv \left( s_W^2Q_q - \frac{1}{2}T_q^3 \right)
c_j^Z  \Pi_Z \pm  Q_q s_W^2c_W^2 \Pi_{\gamma} ~.
\end{equation}

\noindent Here $c_{-1}^Z=1/2-s_W^2, c_{1}^Z=-s_W^2$ (recall that $j = - 1(1)$
for a left(right)  handed electron). $s_W(c_W)$ is the $\sin(\cos)$ of the
weak mixing angle $\theta_W$ and    $Q_q$ and $T_q^3$ are the charge and
$z$-component of the  weak isospin of the quark, respectively. 
Furthermore:

\begin{equation}
\Pi_Z \equiv \left(s-m_Z^2\right)^{-1}~~,~~\Pi_{\gamma}  \equiv s^{-1}~~,~~
\Pi_{q(\bar q)} \equiv \left(m^2_Z + 2p_{q(\bar q)}\cdot p_Z\right)^{-1}~,
%\Pi_{\bar{q}}\equiv \left( m^2_Z + 
%2p_{\bar{q}}\cdot p_Z\right)^{-1} ~,
\end{equation}

\noindent where $s=(p_-+p_+)^2$ is the c.m.\ energy squared of the colliding
electrons,  $s_t\equiv(p_q+p_{\bar{q}})^2$ and we have also defined the CP-odd
quantity  $f\equiv (p_- - p_+)\cdot(p_q+p_{\bar{q}})$. 

As mentioned earlier, being an odd function of the $T_N$ symmetry operation,
$f^1_{--(j)}$ can  only probe CP-asymmetries of the $T_N$-odd type in $e^-e^+
\to t \bar t Z$. This leads us to consider the following dimensionless  
CP-odd, $T_N$-odd observables:

\begin{equation}
O = \frac{\vec{p}_-\cdot(\vec{p}_q\times\vec{p}_{\bar{q}})}{s^{3/2}} 
 ~~,~~ O_{\rm opt} = \frac{\Sigma_{--}^{0({\cal H})}}{\Sigma^0_{+}}
\label{oopt}~.
\end{equation} 

\noindent $O_{\rm opt}$ is an optimal observable in the sense that the statistical
error, in  the measured asymmetry, is minimized \cite{sonioptimal}. Also,
note that both observables are proportional  to $\epsilon(p_-,p_+,p_q,p_{\bar
q})$ since there is only  one possible independent triple correlation product
(or equivalently  a Levi-Civita tensor) when the final state consists of
three particles only and the spins are disregarded. In particular, $O_{\rm
opt}$ is related to $O$  by a multiplication by a CP-even function. The theoretical
statistical significance, $N_{SD}$, in which an   asymmetry can be measured
in an ideal experiment  is given by $N_{SD}= A \sqrt L \sqrt {\sigma_{ttZ}}$
where for the observables $O$ and $O_{\rm opt}$ the  asymmetry $A$, defined
above, is: 

\begin{equation}
A_O \approx \langle O \rangle/\sqrt{\langle O^2\rangle} ~~,~~
A_{\rm opt} \approx \sqrt{\langle O_{\rm opt} \rangle} \label{asym}~.
\end{equation}

\noindent Also, $\sigma_{ttZ} \equiv \sigma(e^-e^+ \to t \bar t Z)$ 
is the cross-section and $L$ is the effective luminosity for fully reconstructed
$t \bar t Z$ events. In particular,  we will take $L=\epsilon {\cal L}$,
where ${\cal L}$ is the total yearly integrated  
luminosity\footnote{For illustrative purposes, we will choose: ${\cal L}=200$
[fb]$^{-1}$ for $\sqrt s=1$ TeV and  ${\cal L}=500$ [fb]$^{-1}$ for $\sqrt
s=1.5$ TeV \cite{nlc}.}  and $\epsilon$ is the overall efficiency for   
reconstruction of the $t \bar t Z$ final state. 
Note that detection of the asymmetries in Eq.~\ref{asym} requires the 
identification of the $t$ and $\bar t$ as well as the reconstruction of their 
momenta. Thus, the most suitable scenario is when either the $t$ or the $\bar
t$ decays  semi-leptonically and the other decays hadronically. Distinguishing
between $t$ and $\bar t$ in the double hadronic decay case will require 
more effort and still remains an experimental challenge.  
%and the 
%factor $\frac{8}{9}$ is added due to the fact that we assume 
%on reconstruction of the $t \bar t Z$ final state 
%when both the $t$ and the $\bar t$ decay leptonically, in 
%which case there will be two missing energies from the two 
%escaped neutrinos.    

In Fig.~2 we plot the cross-section, $\sigma_{ttZ}^{II}$ as a function of
$m_h$ and $\sqrt s$, for Model~II with $\left\{ \tan\beta,\alpha_1,\alpha_2,\alpha_3
\right\}= \left\{ 0.3,\pi/2,\pi/4,0 \right\}$ which we denote as set~II. We will adopt set~II later also when discussing the CP-violating effect.
Here we have set the mass of the heavier Higgs to be $m_H=750$ GeV.  
We see that $\sigma_{tt Z}^{II}$ is typically $\sim {\rm few}~{\rm fb}$ for
c.m. energies of $\sim 1-2$ TeV; it peaks for $m_h \approx 2 m_t$ and  $\sqrt
s \sim 800$ GeV at around 7 fb.\footnote{Plots of the SM cross-section, $\sigma_{ttZ}^{SM}$,
can be found in \cite{hagiwara}, where it was also found that $\sigma_{ttZ}^{SM}
\sim {\rm few}~{\rm fb}$.} Therefore,  with ${\cal L} \gsim 10^2$ [fb]$^{-1}$
it may be possible  to produce $10^2-10^3$ $t \bar t Z$'s at the NLC running 
with c.m. energies $\gsim 1$ TeV scale. We also remark that, although the
cross-sections in the SM and Model~II are of the same order, for certain
values of $\left\{ \tan\beta,\alpha_1,\alpha_2,\alpha_3 \right\}$ and  the
neutral Higgs mass, there can be a significant difference  between the cross-sections
predicted by the two-models. For example, with unpolarized 
incoming electrons and for $\sqrt s = 1$ TeV, $m_h=360$ GeV and $\left\{
\tan\beta,\alpha_1,\alpha_2,\alpha_3 \right\}=\left\{ 0.3,\pi/2,\pi/4,0
\right\}$, $\sigma^{II}_{ttZ} \simeq 6$ fb, while the SM cross-section is
$\sigma^{\rm SM}_{ttZ} \simeq 3.5$ fb.
The  combined information from a study of the cross-section itself along
with CP-violation  may be extremely useful in understanding the
dynamics of the  reaction $e^+e^- \to t \bar t Z$ although we choose not to pursue in that direction in this paper.

Let us first consider an unpolarized incoming electron beam and concentrate
on the CP-odd effect associated with $O_{\rm opt}$. The effect  of the simple
triple product $O$ is slightly smaller.  In Fig.~3 we present our main results
for the expected  asymmetry and statistical significance corresponding to 
$O_{\rm opt}$ in Model~II, as a function of the mass ($m_h$) of the light 
Higgs where, again, $m_H=750$ GeV. We plot $N_{SD}/\sqrt L$,
thus scaling out the luminosity factor from the theoretical prediction and
as an illustration, for the free parameters of Model~II, we adopt set~II defined above, i.e. $\left\{ \tan\beta,\alpha_1,\alpha_2,\alpha_3 \right\}=\left\{
0.3,\pi/2,\pi/4,0 \right\}$. 
We remark that the effect is practically insensitive to $\alpha_3$, and 
$\alpha_1=\pi/2, \alpha_2=\pi/4$ correspond to the best effect, though not
unique.  Also, the CP-violating effect is roughly proportional to $1/\tan\beta$,
it therefore drops as $\tan\beta$ is increased. However, we find that we can
still have $N_{SD}/\sqrt L>0.1$ even in the unpolarized case for $\tan\beta
\lsim 0.6$,  $\alpha_1=\pi/2,\alpha_2=\pi/4,\alpha_3=0$.  

Evidently the asymmetry is almost insensitive to $m_h$
in the range  50 GeV${} \lsim m_h \lsim 2m_t$ where it stays roughly 
at the $7-8\%$ level for $\sqrt s \sim 1-2$ TeV (see Fig.~3). In that range $0.1 \lsim
N_{SD}/\sqrt L \lsim 0.2$; it slightly grows  as $m_h$ is increased and reaches
its peak around  $m_h \approx 2m_t$. For $m_h > 2m_t$, for which an on-shell
$h$ is produced and then decays to a pair of $t \bar t$, as $m_h$ grows
the asymmetry drops till it essentially vanishes when  $m_h \to m_H$ in
which case the ``GIM'' like cancellation,  discussed above, applies. Also,
with respect to the c.m. energy, both $A_{\rm opt}$ and  $N_{SD}/\sqrt L$
reach their peak values  at around $\sqrt s \sim 1$ TeV for both $m_h=100$
and 360 GeV and weakly fall as the c.m.\ energy is increased. For example,
we find that with $m_h=100(360)$ GeV and $\epsilon = 0.5$, it may be possible
to observe a CP-nonconserving effect in  the reaction $e^-e^+ \to t \bar
t Z$ with a statistical  significance of $N_{SD} \approx 1.6(2.0)$ for $\sqrt
s =1$ TeV  and ${\cal L}=200$ [fb]$^{-1}$, and $N_{SD} \approx 2.1(2.5)$
for $\sqrt s =1.5$ TeV  and ${\cal L}=500$ [fb]$^{-1}$ if the incoming electrons
are unpolarized. 

In Table~1 we present $N_{SD}$ for $O_{\rm opt}$, in Model~II with set~II,
for polarized and unpolarized electrons.  For illustrative purposes, we choose
$m_h=100,160$ and 360 GeV and, as before, we present the numbers for $\sqrt s =1$ TeV 
with ${\cal L}=200$ [fb]$^{-1}$ and for $\sqrt s =1.5$ TeV with ${\cal L}=500$
[fb]$^{-1}$. In both cases we take $\epsilon=0.5$ assuming that there is
no loss  of luminosity when the electrons are polarized.\footnote{If the efficiency for $t \bar tZ$ reconstruction is $\epsilon=0.25$ then our numbers would correspondingly require 2 years of running.}
Also, to demonstrate the sensitivity of the CP-effect to the mass ($m_H$) of the heavy Higgs, we present numbers for both $m_H=750$ GeV (shown in the parentheses) and $m_H=1$ TeV. We see from Table~1
that left polarized incoming electrons can probe CP-violation slightly better
than  unpolarized ones. In particular, for $m_H=1$ TeV we find that with left polarized electrons
and for $\sqrt s =1(1.5)$ TeV,  the CP-effect is above the 2(3)-sigma level
%ASf ..remove extra GeV
for  $m_h \gsim 2m_t$. At c.m. energy $\sqrt s=1$ TeV 
the CP-effect is practically insensitive to the choice $m_H=750$ GeV or $m_H=1$ TeV. However, we see from the Table that as one goes to $\sqrt s=1.5$ TeV, $m_H=1$ TeV can give rise to a 3-sigma signal in the 
range, $100 ~{\rm GeV} \lsim m_h \lsim  400 ~{\rm GeV}$, 
if the electrons are negatively polarized, while with $m_H=750$ GeV 
the CP-signal is smaller by about half a sigma. 
We remark again that the results for the simple observable
%ASf remove extra ,  
$O$ exhibit the same  behavior though slightly smaller then those for $O_{\rm
opt}$.  
%ASf in the previous para changed table to Table and changed the m_h inequality

Before summarizing we wish to emphasize that the analysis performed here
for Model~II can be generalized  to models I and III as well. Recall that in Model~II a mass-dimension-two operator, that softly breaks the discrete symmetry which is responsible for natural-flavor-conservation (NFC), is needed in order to have CP-violati
ng Higgs-fermion couplings. On the other hand, 
in Model~III there is no NFC and the CP-odd phase in the ${\cal H}^k qq$ vertex can arise
from a phase in the Yukawa couplings  $U^2_{ij}$ and $D^2_{ij}$ defined in
Eq.~\ref{yukawa}  (for more details see \cite{sonirev}). The pseudoscalar 
coupling in Eq.~\ref{hqqhzz}, responsible for CP-violation,  can be chosen
(in Model~III) as $b^k_q \propto \lambda_q$ \cite{sher}, where $\lambda_q$ is a free
parameter of the model expected to be of ${\cal O}(1)$ \cite{sonirev}.
Then the replacement  $\lambda_t \approx 1/\tan\beta$ (for a given value
of $\tan\beta$ in Model~II) can give rise to  comparable CP-nonconserving
effects in $e^+e^- \to  t \bar t Z$. Thus, the main difference between 
Model~II and Model~III arises from the fact that, while in Model~II a small 
$\tan\beta$ is required in order to enhance the CP-odd effect, in Model~III
the effect is elevated as $\lambda_t$ is correspondingly increased. The same argument holds
also for other previously suggested  CP studies in top systems where the
results obtained for Model II can be generalized to Model III\null. This  
seems to indicate that even if such CP-violating effects are found in $e^-e^+
\to t \bar t Z$ or other reactions which involve the top-neutral Higgs CP-phase
discussed in this paper, it may not be possible  to distinguish Model~II
from  Model~III, as both models may have a comparable CP-odd phase in the
Higgs sector. In that sense,  the best way to proceed for correctly classifying
the Higgs sector,   is to search for large signatures of FC effects in top
quark reactions as well. Model~III, with FCS couplings to fermions proportional
to the fermion masses involved in the FC vertex \cite{sher}, may indeed drive
such large FC  effects in top systems, some of which are $e^+e^- \to t \bar
c;~t \bar c \nu_e {\bar {\nu}}_e;~ t \bar c e^+e^-;~t \bar c Z;~ t \bar t
c \bar c;~t \bar c q \bar q$ and were  investigated in \cite{fc2hdm}. Detection
or no detection of these  FC signatures along with evidence for CP-violation
in the Higgs  sector in high energy $e^+e^-$ colliders, may well be the only
way  to experimentally distinguish between scalar dynamics of a Model~II 
or a Model~III origin.      

To summarize, CP-violation in $e^-e^+ \to t \bar t Z$ at a future high energy
$e^+e^-$ collider was studied in the context of 2HDM's. An important property of this
reaction is that  CP-violation arises already at the tree-level through interference
of $Z$ emission from the $t$ or $\bar t$ and its emission off a s-channel
$Z$ and therefore allowing for a relatively large CP-violating  signal. In
particular, we found that within a broad range of the lightest Higgs mass,
100 GeV${} \lsim m_h \lsim 400$ GeV and with c.m.\ energies between
1--2 TeV  the asymmetry can reach the $\sim 10\%$ level. The corresponding
statistical significance, in which it may be observed, is around 2--3 sigma
for unpolarized incoming electrons and, if $m_h \gsim 2m_t$, it can reach above 3-sigma for polarized ones and c.m. energy of $\sqrt s= 1.5$ TeV. Bearing the expected difficulty
of observing the Higgs exchange effect in $e^+e^- \to t \bar t Z$ in
the SM and with Higgs masses below $2m_t$ \cite{hagiwara}, it is especially
gratifying that the CP-effect is sizable and  almost insensitive to $m_h$
in the range  50 GeV${} \lsim m_h \lsim 2m_t$  GeV\null. We therefore
encourage a detailed scrutiny of the reaction  $e^-e^+ \to t \bar t Z$ in
the NLC\null. Such an investigation, especially due to the promising CP-nonconserving
effects reported  here, may be helpful in unraveling the CP properties of the Higgs sector.    

\bigskip  

This research was supported in part by the U.S. DOE contract numbers DE-AC02-76CH00016(BNL), DC-AC05-84ER40150(Jefferson Lab) and DE-FG03-94ER40837(UCR).

\pagebreak

\begin{center}
{\bf Figure Captions}
\end{center}

\begin{description}

\item{Fig. 1:} Tree-level Feynman diagrams contributing to $e^+e^-\to t\bar{t}Z$
in a two Higgs doublet model. Diagram $a$  represents 8 diagrams in which
either $Z$ or $\gamma$ are exchanged in the s-channel and the outgoing $Z$
is emitted from $e^+,e^-,t$ or $\bar t$ 

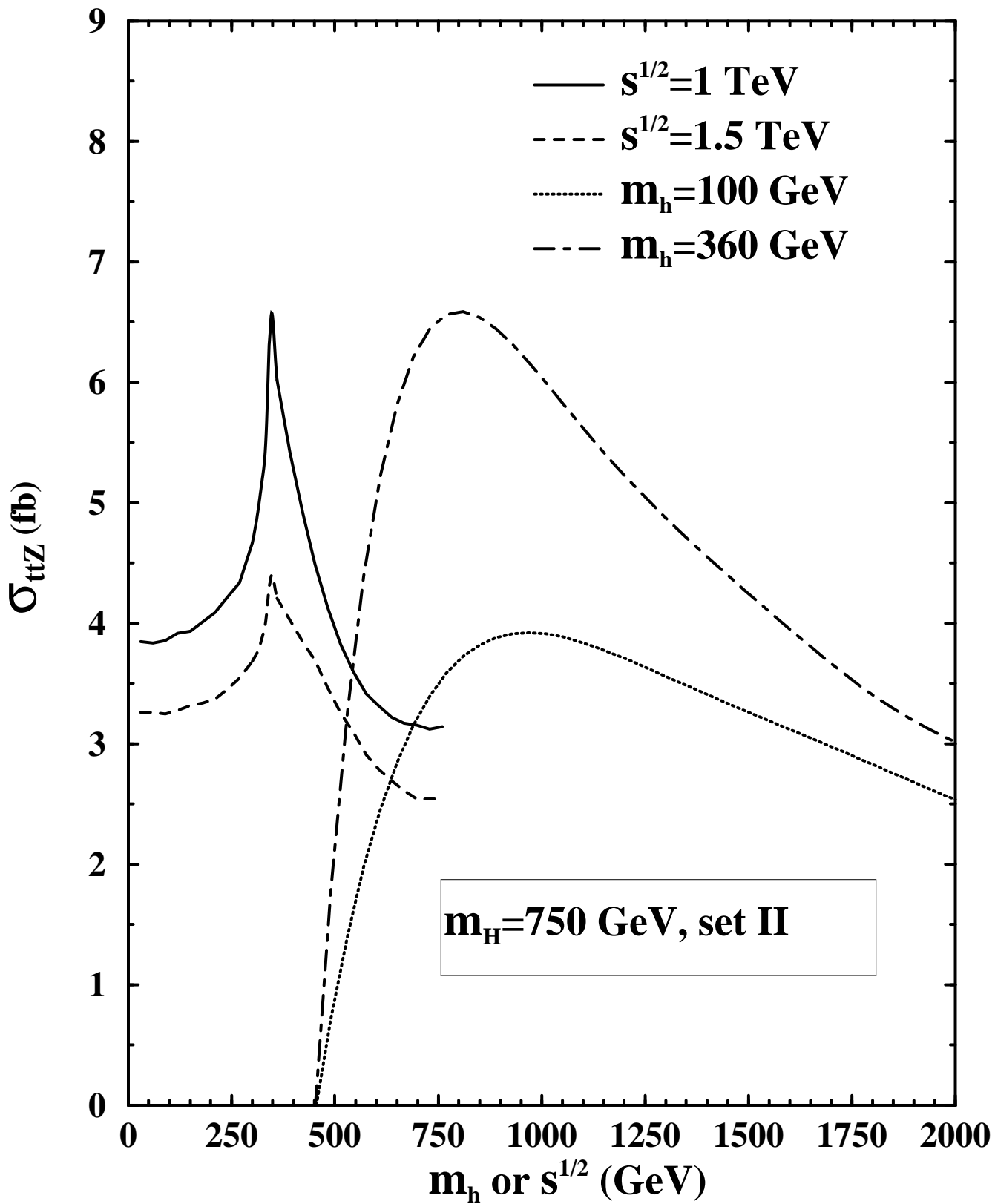
\item{Fig. 2:} The cross section (in [fb]) for the reaction $e^+e^-\to t\bar{t}Z$,
assuming unpolarized  electron and positron beams, for Model~II with set~II
and as a function of $m_h$ (solid and dashed lines) and $\sqrt s$ (dotted
and dotted-dashed lines). Set~II means $\left\{ \tan\beta,\alpha_1,\alpha_2,\alpha_3
\right\} \equiv \left\{ 0.3,\pi/2,\pi/4,0 \right\}$.

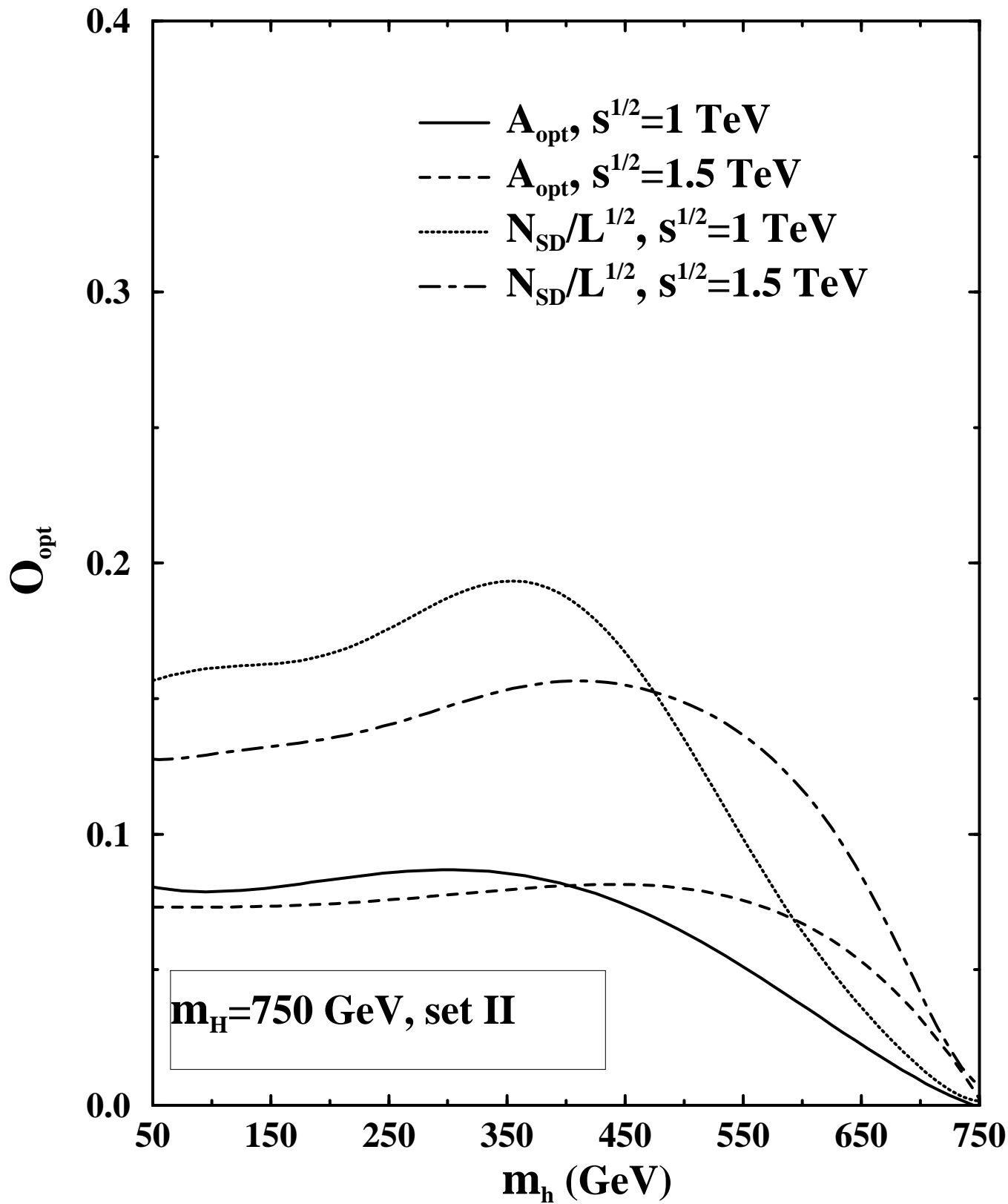
\item{Fig. 3:} The asymmetry, $A_{\rm opt}$, and scaled statistical 
significance, $N_{SD}/\sqrt L$, for the optimal observable 
$O_{\rm opt}$ as a function of the light Higgs mass $m_h$, for
$\sqrt s=1$ TeV and 1.5 TeV\null. See also caption to Fig.~2.

\end{description}

\newpage

\begin{table}
\begin{center}
\caption[first entry]{The statistical significance, $N_{SD}$, in which 
the CP-nonconserving effects in $e^+e^- \to t \bar t Z$ can be detected in
%ASf...extra a removed
one year of running of a future high energy  collider with either unpolarized
or polarized incoming electron beam.  We have used a yearly integrated luminosity
of ${\cal L}=200$ and 500 [fb]$^{-1}$ for $\sqrt s=1$ and 1.5 TeV, respectively,
and an efficiency reconstruction factor of  $\epsilon=0.5$ for both energies. $N_{SD}$ is given for both $m_H=750$ GeV (in parentheses) and $m_H=1$ TeV. 
Recall that $j=1(-1)$ stands  for right(left) polarized electrons. Set~II means $\left\{ \tan\beta,\alpha_1,\alpha_2,\alpha_3
\right\} \equiv \left\{ 0.3,\pi/2,\pi/4,0 \right\}$.

\bigskip

\protect\label{oropttab}}
\begin{tabular}{|r||r|r|r|r|} \cline{3-5}
\multicolumn{2}{c||}{~~} & \multicolumn{3}{c|}{$e^+e^- \to t \bar t Z$ (Model~II with Set~II)}\\ \hline
$\sqrt{s}$ & j
& \multicolumn{3}{c|}{$O_{\rm opt}$}
\\ \cline{3-5}
$({\rm TeV}) \Downarrow$ &$({\rm GeV}) \Rightarrow$ & $m_h=100$ & $m_h=160$ &$m_h=360$\\ 
\hline
\hline
&-1& $(1.8)~1.7$ & $(1.8)~1.8$ & (2.2)~2.2 \\ \cline{2-5}
1 & unpol & $(1.6)~1.6$ & $(1.7)~1.6$ & (2.0)~2.0 \\ \cline{2-5}
& 1 &  $(1.5)~1.5$ & $(1.5)~1.5$ & (1.8)~1.8 \\ \hline
\hline
& -1 &  $(2.3)~2.9$ & $(2.4)~3.0$ & (2.8)~3.3 \\ \cline{2-5}
1.5 & unpol & $(2.1)~2.6$ & $(2.1)~2.7$ & (2.5)~3.0 \\ \cline{2-5}
&1 &  $(1.8)~2.3$ & $(1.8)~2.3$ & (2.1)~2.6\\ \hline 
\end{tabular}
\end{center}
\end{table}

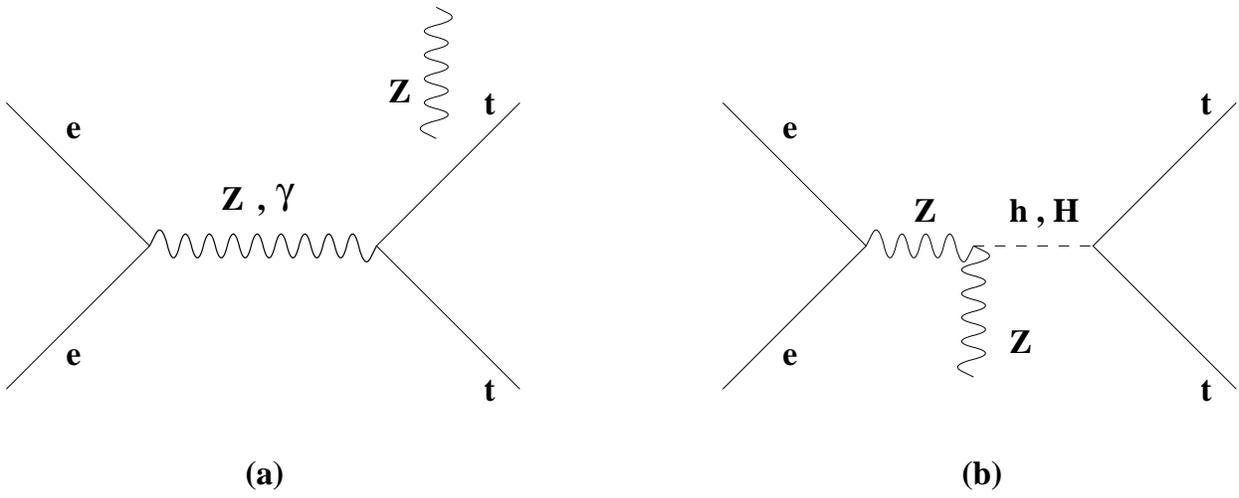
\end{document}